\documentclass[aps,prl,showpacs,twocolumn]{revtex4-1}
\usepackage{fullpage,float,graphicx,mathrsfs,subfig}
\usepackage{amsmath,amssymb,latexsym,float}
\usepackage{tikz}
\usetikzlibrary{calc}
\usetikzlibrary{decorations.pathmorphing}

\newcommand{\bpr}{\begin{proof}}
\newcommand{\epr}{\end{proof}}
\newcommand{\bsol}{\begin{proof}[Solution:]}
\newcommand{\ben}{\begin{enumerate}}
\newcommand{\een}{\end{enumerate}}
\newcommand{\bit}{\begin{itemize}}
\newcommand{\eit}{\end{itemize}}

\begin{document}

\title{Breathers In Periodic Granular Chains With Multiple Band Gaps}
\author{C. Hoogeboom}
\affiliation{Department of Mathematics and Statistics, University of Massachusetts, Amherst MA 01003-4515, USA}
\author{P.G. Kevrekidis}
\affiliation{Department of Mathematics and Statistics, University of Massachusetts, Amherst MA 01003-4515, USA}
\date{\today}

\begin{abstract}
	We consider the localized nonlinear breathing modes that emerge in heterogeneous granular configurations of two materials with a periodicity of three and four beads. 
	We examine as characteristic examples chains with one steel and two aluminum beads, as well as ones with 1 steel and three aluminum beads, although we briefly touch upon other configurations as well, illustrating their similar characteristics. 
	We analyze the higher order gaps that emerge in such settings and explore the intrinsic localized modes that bifurcate from the edge of the upper bands. 
	A generic surprising feature of such states is that they appear to be more robust than their counterparts bifurcating from the edges of the lower bands. 
	Direct numerical simulations, using driving of the system at suitable frequencies through an actuator or taking advantage of the modulational instabilities of extended band edge states in the system illustrate the spontaneous formation of localized modes within the corresponding nearest gaps. 
	In these numerical experiments, we also account for the presence of dissipation and  analyze its potential role.
\end{abstract}

\maketitle

\section{Introduction}

	The theme of ``granular crystals,'' i.e., of chains composed of elastic (typically spherical) particles in close contact with one another, is one that has been gaining considerable momentum over the past few years \cite{nesterenko1,sen08}. 
	Of particular interest in this setting is the nonlinear interaction between the grains, which features a so-called Hertzian contact between the adjacent beads in the chain upon compression and a zero tensile strength. 
	The Hertzian interaction potential depends on the relative displacement of the bead centers raised to the $5/2$ power for spherical beads. 
	One of the remarkable features of this system is that it can be tuned to possess a dynamic response ranging (depending on the static compression applied to the chain) from weakly all the way to strongly nonlinear \cite{nesterenko1, dar06}.
	It is precisely this tunability and straightforward experimental implementation of such chains that enables a wide range of nonlinear dynamical studies in them, featuring the interplay of discreteness, nonlinearity and periodicity (or lack thereof). 
	Additionally, such systems have shown considerable potential for a wide range of engineering applications including shock and energy absorbing layers \cite{dar06,hong05,fernando,doney06}, actuating devices \cite{dev08}, acoustic lenses \cite{Spadoni}, acoustic diodes \cite{china}, and sound scramblers \cite{dar05,dar05b}.

	Our emphasis in the present work will be on the coherent nonlinear waveforms that can trap the energy in such systems. 
	Among the most prominent ones such that have been reported in the literature are the solitary waves \cite{nesterenko1,Coste97,sen08,pikovsky}, shock waves \cite{chiara09,vakakis10} and very recently discrete breathers \cite{Boechler10,theo10,hoogeboom} (see also related discussions in Refs. \cite{mohan,job}). 
	Our emphasis in the present work will be on the discrete breathers, which have a time honored history in a diverse host of settings. 
	Such intrinsic localized modes constitute generic excitations of nonlinear dynamical lattices, which are (typically) exponentially localized in space, while they oscillate periodically in time. 
	The discrete breathers appear to be relevant to a wide array of thematic areas spanning superconducting Josephson junctions~\cite{orlando1}, photonic crystals~\cite{photon}, biopolymers~\cite{Xiepeyrard}, charge-transfer solids~\cite{Swanson1999}, antiferromagnets~\cite{Schwarz1999}, and micromechanical cantilever arrays~\cite{sievers}, among others~\cite{flach09}.
	
	It has been argued \cite{Boechler10,theo10} that  discrete breathers are likely to be absent in monoatomic granular chains~\cite{chong}.
	Hence, their first theoretical investigations and experimental implementations in the above works took place in the context of heterogeneous systems in the form of {\it dimers}. 
	Such arrays consisting of glued \cite{Hennion07}, welded \cite{Hennion05}, and elastically compressed spherical particles \cite{Billy00,Boechler09,Boechler10,Herbold} have been shown to exhibit tunable frequency vibrational band gaps. 
	The works of Refs. \cite{Boechler10,theo10} took advantage of the modulational instability \cite{flach09} of the mode at the bottom edge of the optical band in this type of chain to produce dynamical localization in the form of discrete breathers whose frequency of oscillation resided in the nearest bandgap (while the length scale of their exponential localization depended on how deep in the bandgap their frequency was lying).

	However, in recent experimental works \cite{mason1,mason2,dimerlin}, experimental accessibility has been demonstrated towards states with generalized periodicities that are not restricted merely to dimers, but may include {\it trimers} (such as so-called $2:1$ configurations, with $2$ beads of one material and one bead of another, or $1:1:1$ chains consisting of $3$ different material types) or even {\it quadrimers} (with $3:1$, $1:1:1:1$, $2:2$, $2:1:1$ constituting possible combinations of materials therein). 
	While solitary traveling waves have been identified in such chains in the work of Refs. \cite{mason1,mason2}, and their linear spectrum under precompression has been illustrated to contain multiple bandgaps in the recent work of Ref. \cite{dimerlin}, nonlinear breathing excitations have not, to the best of our knowledge, been identified therein. 
	This constitutes the focus of the present study, which is partly motivated from the analog of such investigations in nonlinear Schr{\"o}dinger models with optical lattices (related to Bose-Einstein condensates) whereby gap solitons in each of the gaps of the optical lattice potential have been identified in the recent work of Ref. \cite{biaowu}.

	Herein, upon presenting the general theoretical setup (Sec. 2), we restrict ourselves to arguably the simplest trimer and quadrimer configurations, namely, the experimentally accessible $2:1$ chains with two aluminum and one steel bead (Sec. 3), as well as their $3:1$ counterparts with three aluminum and one steel bead (in Sec. 4). 
	Nevertheless, the generality of our conclusions is shown to be supported by the examination of additional configurations, such as two steel beads and one aluminum, or a trimer that involves aluminum, steel, and brass.

	We commence their examination by analyzing the corresponding linear spectrum. 
	We find here (as well as more generally) that configurations with a periodicity of $P$ beads in such granular systems possess $P-1$ finite gaps and a semi-infinite one (extending from the top of the uppermost optical band to $\infty$). 
	Within the resulting 2 finite gaps of the trimer chain and the 3 of the quadrimer one we seek to identify discrete breather states and systematically continue them within the relevant bandgap. 
	We believe that this is the first demonstration of such higher bandgap states in granular systems. 
	Moreover, such states have a rather surprising characteristic: They appear to be considerably more robust than their lower gap counterparts; that is, they have wider parametric intervals of stability and even when unstable their instability growth rates are typically considerably lower than their lower bandgap counterparts. 
	Finally, in each case, to demonstrate the experimental realizability of the pertinent states, we report dynamical simulations under experimentally realistic conditions; in that light we also account for the presence of dissipation and elucidate its potential role. 
	We induce the formation of these nonlinear gap solitary wave (or gap breather) states through the excitation of a modulationally unstable mode of the system at the frequency of the bottom of the optical band edge (different periodicities exhibit different numbers of such band edge modes). 
	This is either done by initializing the system at the spatial profile of the  extended mode of the proper frequency, or, more realistically, by driving (``actuating'') the system at the relevant frequency; cf. also the experimental realization of the lowest bandgap breathers in Ref. \cite{Boechler10}. 
	Finally, in Sec. 5, we summarize our findings and present our conclusions.

%%%%%%%%%%%%%%%%%%%%%%%%%%%%%%%%%%%
% Section %   Theoretical Setup   %
%%%%%%%%%%%%%%%%%%%%%%%%%%%%%%%%%%%
%
\section{Theoretical Setup}
%
%
%%%%%%%%%%%%%%%%%%%%%%%%%%%%%%%%%%%
% Figure %       Schematic        %
%%%%%%%%%%%%%%%%%%%%%%%%%%%%%%%%%%%
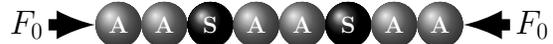
\begin{figure}
\resizebox{.45\textwidth}{!}{
\begin{tikzpicture}
	[circleball/.style={
		shape=circle, 
		minimum size=1cm, 
		shade, 
		shading=ball,
		text=white	}]
\node (f01) at (-2,0) {{\huge $F_0$}};
\def\numberOfBalls{6}
\foreach \x in {0,3,...,\numberOfBalls}{
	\node[circleball, ball color=black!50] (ball0\x)at (\x,0) {{\Large \bf A}};
	\node[circleball, ball color=black!50] (ball1\x)at ($(\x+1,0)$) {{\Large \bf A}};
	\ifnum \x<\numberOfBalls
		\node[circleball, ball color=black] at ($(\x+2,0)$) {{\Large \bf S}};
	\fi
}
\node (f02) at ($(\numberOfBalls+3,0)$) {{\huge $F_0$}};

\path[line width=7pt] (f01) edge[-latex] (ball00)
					  (f02) edge[-latex] (ball1\numberOfBalls);

\end{tikzpicture} }
\caption{
	(Color online) A schematic of our $2:1$ aluminum-steel configuration.
}\label{f:schematic}
\end{figure}
%%%%%%%%%%%%%%%%%%%%%%%%%%%%%%%%%%
%
%
	We will be focusing on a one-dimensional (1D) granular crystal containing steel and aluminum spheres compressed uniaxially by a force $F_0 = 22.4 N$ (this ensures the existence of a linear spectrum, which is critical for enabling the generation of breathers within the spectral gaps). 
	This granular crystal is modeled by a system of coupled nonlinear oscillators; the equation of motion for the displacement $u_n$ of the $n$th bead is given by \cite{nesterenko1,sen08}
\begin{eqnarray} % Equations of motion
m_n\ddot{u}_n &=&A_{n-1,n}[\delta_{n-1,n}-(u_n-u_{n-1})]_+^{3/2}
\nonumber 
\\
&-& A_{n,n+1}[\delta_{n,n+1}-(u_{n+1}-u_n)]_+^{3/2},\label{e:motion}
\end{eqnarray}
where $m_n$ is the mass of the $n$th bead, $A_{n,n+1}$ is the coefficient of elasticity between beads $n$ and $n+1$, and $\delta_{n,n+1}=(F_0/A_{n,n+1})^{2/3}$ is the amount of static overlap of the beads $n$ and $n+1$ (due to the precompression) under equilibrium conditions.  
	The coefficient $A_{n,n+1}$ depends on the material and geometric properties of the beads $n$ and $n+1$; for sphere-to-sphere contact, which will be the case of interest herein, the coefficient $A_{n,n+1}$ is given by
\begin{equation} % Elastic Coefficient
A_{n,n+1}=\frac{4 E_{n}E_{n+1}\sqrt{\dfrac{r_nr_{n+1}}{r_n+r_{n+1}}}}{3E_{n+1}(1-\nu_{n}^2)+3E_n(1-\nu_{n+1}^2)}\label{e:forcecoeff}
\end{equation}
where $r_n$, $E_n$, and $\nu_n$ denote, respectively, the radius, elastic (Young's) modulus, and Poisson ratio of the $n$th bead.
	The bracket $[s]_+$ of Eq.~(\ref{e:motion}) is the positive part of $s$, that is,
$$
[s]_+=\max(0,s)
$$ 
	This accounts for the fact that if two beads are not in contact, they do not exert any force on each other (zero-tension).
	In addition, each bead (away from the boundary) has a local energy density associated with it, given by
\begin{equation} % Energy 
e_n=\frac{1}{2}m_n\dot{u}_n^2+\frac{1}{2}[V_{n-1}(u_n-u_{n-1})+V_n(u_{n+1}-u_n)],\label{e:energy}
\end{equation}
where the interaction potential $V_n(u_{n+1}-u_n)$ is given by
\begin{align} % potential
V_n(u_{n+1}-u_n) = &\frac{2}{5}A_{n,n+1}[\Delta_{n,n+1}-(u_{n+1}-u_n)]_+^{5/2}\nonumber\\
& - A_{n,n+1}\Delta_{n,n+1}^{3/2}(u_{n+1}-u_n)\nonumber\\
& - \frac{2}{5}A_{n,n+1}\Delta_{n,n+1}^{5/2}. \label{e:potential}
\end{align}
	We note here that the sum of \eqref{e:energy} over all the beads gives us the Hamiltonian of the system, and the equations of motion can then be derived in the usual way (using Hamilton's equations).
	In Table \ref{t:matprops}, we present an overview of the material properties used, in sync with earlier experimental results \cite{Boechler10}.
	The trimer lattice will have repeating unit cells containing a `aluminum-aluminum-steel' (a-a-s) pattern; see Fig. \ref{f:schematic}.
	We will also consider chains with four beads per unit cell, i.e., an `aluminum-aluminum-aluminum-steel' (a-a-a-s) pattern. 
	Although additional heterogeneous configurations are possible, per our discussion in the previous section, our results indicate that these settings accurately reflect the potential for classes of additional gap breather modes.

\begin{table} % Material Properties
\begin{tabular}{l  r r r}
\hline\noalign{\smallskip}
Material & Steel & Aluminum\\[3pt]
\hline\noalign{\smallskip}
Radius (mm) &9.53 &9.53 \\[3pt]
\hline\noalign{\smallskip}
Density ($\text{kg}/\text{m}^3$) &8027.17 &2700\\[3pt]
\hline\noalign{\smallskip}
Elastic modulus (Pa) &$2\times 10^{11}$ &$7\times 10^{10}$ \\[3pt]
\hline\noalign{\smallskip}
Poisson ratio &0.3 &0.35\\[3pt]
\hline
\end{tabular}
\caption{
	The material properties used in our simulations; cf. the experimental parameters used in Ref. \cite{Boechler10}.
}\label{t:matprops} 
\end{table}

%%%%%%%%%%%%%%%%%%%%%%%%%%%%%%%%
% Section %    A-A-S Chain     %
%%%%%%%%%%%%%%%%%%%%%%%%%%%%%%%%
\section{A-A-S chain}
%
%
%
%
%%%%%%%%%%%%%%%%%%%%%%%%%%%%%%%%
%          %  Linear Spectrum  %
%%%%%%%%%%%%%%%%%%%%%%%%%%%%%%%%
\subsection{Linear Spectrum}
%
%
%
%%%%%%%%%%%%%%%%%%%%%%%%%%%%%%%%
% Figure %   Linear Spectrum   %
%%%%%%%%%%%%%%%%%%%%%%%%%%%%%%%%
%
\begin{figure}
\includegraphics[width=.45\textwidth]{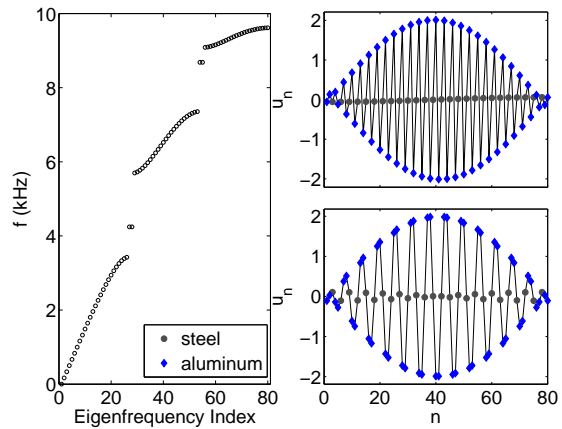}
\caption{
	(Color online) \emph{Left:} The linear spectrum for a chain of 80 beads with period 3 (aluminum-aluminum-steel).
	There are three linear bands separated by two finite gaps. 
	\emph{Right:} The eigenmodes from the bottom edges of (\emph{top to bottom}) the second and first optical bands of the chain.}
\label{f:aas_linear_spectrum}
\end{figure}
%%%%%%%%%%%%%%%%%%%%%%%%%%%%%%%%
%%%%%%%%%%%%%%%%%%%%%%%%%%%%%%%%
%
%
%
	Except for some degenerate (non-generic) cases, which depend on the materials used, the periodicity of the lattice determines the number of distinct bands in the dispersion relation; each band is a family of solutions to a polynomial equation (specifically a $P\times P$ matrix eigenvalue problem), and the degree of the polynomial is the same as the period of the lattice.
	Thus, in general, the number of finite gaps in the spectrum is one less than the period of the lattice (there is generically also a semi-infinite gap extending from the top of the highest optical band to $+\infty$). 
	Figure \ref{f:aas_linear_spectrum} shows the spectrum for an a-a-s lattice along with sample eigenmodes from the bottom edges of the bands; these will be used as the initial seeds for our nonlinear solvers that will be used to identify the discrete breather modes in the gaps of the corresponding linear spectra.

	Some of the features worth mentioning within the relevant spectrum are the following. 
	Firstly, we can observe that the free boundary conditions used, along with lighter masses at each end of the chain, induce the presence of some localized modes centered around the edge of the chain with frequencies within the linear gap. 
	Such ``surface modes'' have been discussed earlier in the context of Refs. \cite{theo10,hoogeboom} and will not be further elaborated herein. 
	The modes at the bottom edges of both optical bands are reminiscent of the ones reported in the dimer case in Ref. \cite{Boechler10} in that the ``heavy'' steel beads are immobile while the ``lighter'' aluminum ones are the ones oscillating. 
	As may be anticipated, in the lower band, these aluminum pairs oscillate in phase, while in the upper one, they oscillate out of phase. 
	It should be noted that a detailed analysis (through explicit dispersion relations) of the linear spectrum in the case of an a-a-s lattice can be found in the recent theoretical and experimental work of Ref. \cite{dimerlin}; hence we refer the interested reader there for further details on the linear picture. 
%
%
%
%%%%%%%%%%%%%%%%%%%%%%%%%%%%%%%%%%%
%           % Existence/Stability %
%%%%%%%%%%%%%%%%%%%%%%%%%%%%%%%%%%%
\subsection{Existence and Stability of Discrete Breathers}
\begin{figure}
\includegraphics[width=.45\textwidth]{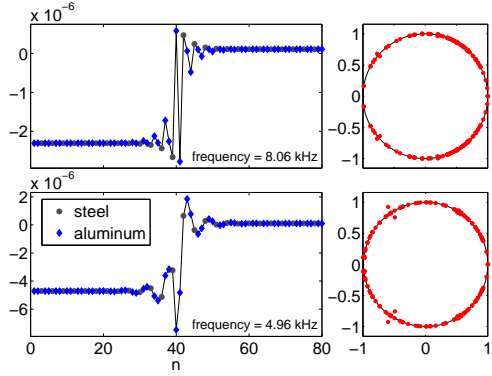}
\caption{
	(Color online) \emph{Top:} A sample breather profile from the upper band gap, along with its Floquet multipliers.
	\emph{Bottom:} A sample breather from the lower band gap, along with its Floquet multipliers.
}
\label{f:aas_breather_profiles}
\end{figure}

	Using an initial guess from the bottom of the band edge, we performed a numerical single-parameter continuation of a Poincar\'{e} map through each spectral gap using frequency as our continuation parameter.
	See the Appendix for more details about this method of continuation, and the Floquet multipliers.
	Sample breather profiles from these continuations are shown in Fig. \ref{f:aas_breather_profiles}.
	It can be observed that similarly to the profiles of the linear modes from which they stem, the breather modes bear an apparent in-phase excitation of their aluminum beads in the lower gap, and an out-of-phase one in the upper gap. 
	In these nonlinear modes, the heavier steel beads gradually become excited as well. 
	Furthermore, the deeper into the respective gaps the modes get (in terms of frequency), the more localized they become, as is evident by the strongly localized
form of the gap breather profiles for the chosen frequencies lying around the middle of the respective gaps.

\begin{figure}
\includegraphics[width=.45\textwidth]{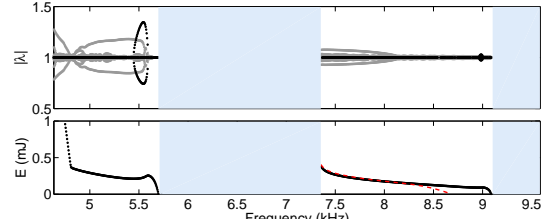}
\caption{
(Color online) \emph{Top:} The magnitude of each Floquet multiplier is shown as a function of its frequency.
	Purely real Floquet multipliers are shown in black, while Floquet multipliers associated  with nonzero imaginary part are shown in gray.
	\emph{Bottom:} The energy of each solution as a function of frequency. 
	The sharp corner in the energy happens at the critical frequency $f_c$, where the breathers start resonating with the linear modes due to the second harmonic.
	The horizontal scale is the same for both panels, and the frequency ranges of the linear bands are shaded for clarity.
}
\label{f:aas_energy_floq}
\end{figure}

	In parallel to the numerical continuation, we also examined the  linear stability of each obtained breather solution via Floquet analysis. 
	The relevant results are shown in the full linearization 
spectra (corresponding to the breather
profiles) of~Fig. \ref{f:aas_breather_profiles} and the 
continuation results of~Fig. \ref{f:aas_energy_floq}. 
	Whenever any of these Floquet multipliers $\lambda_i$ satisfies $|\lambda_i|>1$, the corresponding breather is linearly unstable. 
	There are two types of instabilities that we observe in this Hamiltonian system. 
	The one that stems from a purely real Floquet multiplier with magnitude greater than one (which forms a pair with another such with magnitude less than one), we will call a real instability. 
	On the other hand, there also exists the possibility of a quartet of Floquet multipliers outside the unit circle. 
	These modes bearing a nonzero imaginary part are typically weaker, and we will call them oscillatory instabilities, as they are associated with complex eigenvalues and exhibit both oscillatory and growth dynamics. 
	
	As has been observed previously in Refs. \cite{Boechler10,theo10,hoogeboom}, a change in the monotonicity of the energy as a function of frequency is directly associated with a real instability for the case of gap breathers.
	Such an instability was earlier observed for the first gap in the dimer case in Refs. \cite{Boechler10,theo10}, and a similar finding arises in the lower finite gap of the trimer case considered herein. 
	Furthermore, the structure of the oscillatory instabilities appears to be similar between the two settings. 
	Interestingly, once the frequency becomes lower than a critical point, call it $f_c$, sharp resonances with extended modes emerge similar to what was observed in Ref. \cite{theo10} and the gap breathers become highly unstable (hence, we do not consider such modes for lower frequencies herein).
	This critical point $f_c$, as it turns out, is the point at which the 2nd harmonic of the breather frequency lies in the highest linear band, and thus the breather is resonating with the linear modes at the second harmonic.
	This is confirmed by observing that the breathers found with frequency lower than $f_c$ have excited tails, which oscillate with a frequency twice that of the breather.

	However, the key observation of the present work is that the modes of the higher gap have a significantly narrower frequency range for which they exhibit the real instability (in fact, for different material properties the instability may be entirely absent; see below).
	Furthermore, for the oscillatory instabilities too, the growth rates appear to be considerably weaker than their lower gap analogs in Fig. \ref{f:aas_energy_floq}, apparently leading to considerably more robust higher gap states. 
	This is a rather surprising observation and may perhaps be attributed to the more pronounced out-of-phase structure of the higher gap modes, which is well known in other classes of dynamical lattice models to offer enhanced stability as well \cite{DNLS}.
	
\begin{figure}
\includegraphics[width=.45\textwidth]{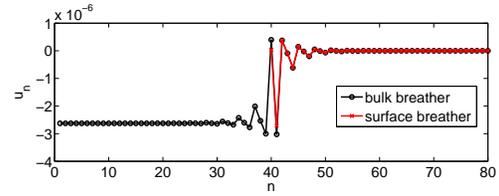}
\caption{
	(Color online) The surface breather with frequency of 7.9764 kHz is shown 
shifted and overlaid on the bulk breather of the same frequency.
}
\label{f:aas_bulk_surface_comparison}
\end{figure}

	For completeness, we also performed a continuation through the spectral gap, starting with an initial seed of the surface mode in the gap, which is present because of the free boundary conditions, and the light masses on the ends.
	These surface breathers were incredibly robust along the entire continuation, with all Floquet multipliers staying within a 0.01 radius of the unit circle.
	The energy of these breathers increased until it closes in on the branch from the bulk of the chain, and then follows closely that branch through the rest of the continuation.
	The profile of the surface breather corresponds almost exactly to the profile of the corresponding bulk breather, as shown in Fig. \ref{f:aas_bulk_surface_comparison}.
	This suggests that surface breathers are effectively bulk breathers which have been shifted to the boundary.

%
%
%
%%%%%%%%%%%%%%%%%%%%%%%%%%%%%%%%%%
% Figure %   Breather Profiles   % 
%%%%%%%%%%%%%%%%%%%%%%%%%%%%%%%%%%
%%%%%%%%%%%%%%%%%%%%%%%%%%%%%%%%%%
%%%%%%%%%%%%%%%%%%%%%%%%%%%%%%%%%%
%
%
%
%%%%%%%%%%%%%%%%%%%%%%%%%%%%%%%%%%%
% Figure %     Energy/Floquet     %
%%%%%%%%%%%%%%%%%%%%%%%%%%%%%%%%%%%
%%%%%%%%%%%%%%%%%%%%%%%%%%%%%%%%%%%%
%%%%%%%%%%%%%%%%%%%%%%%%%%%%%%%%%%%%
%
%
%%%%%%%%%%%%%%%%%%%%%%%%%%%%%%%%%%%%%%%%%%%%%
% Figure % Bulk/Surface Comparison          %
%%%%%%%%%%%%%%%%%%%%%%%%%%%%%%%%%%%%%%%%%%%%%
%%%%%%%%%%%%%%%%%%%%%%%%%%%%%%%%%%%%%%%%%%%%
%%%%%%%%%%%%%%%%%%%%%%%%%%%%%%%%%%%%%%%%%%%%
%
%
%
%%%%%%%%%%%%%%%%%%%%%%%%%%%%%%%%%%%%%
%          %       Dynamics         %
%%%%%%%%%%%%%%%%%%%%%%%%%%%%%%%%%%%%%
\subsection{Dynamics and Modulational Instability}
	We now turn to the examination of the potential experimental feasibility of exciting such nonlinear localized modes in the 2:1 chain. 
	To that effect, we will examine a number of direct numerical simulations of the chain evolution under suitable initial conditions or external drive.	
	In what follows, we focus, in particular, on the breathers from the upper band gap, since they constitute the most robust (and, hence, most likely to be experimentally observable) structures of the system.

\begin{figure}
\includegraphics[width=.45\textwidth]{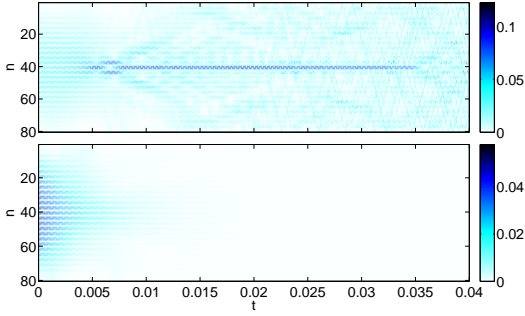}
\caption{
(Color online) \emph{Top:} Modulational instability of the first eigenmode of the upper most band, in a chain with 'aluminum - aluminum - steel' cells. A robust, 
long-lived breather is formed.
\emph{Bottom:} The same as the top panel, except in this case, dissipation is turned on, with $\tau=2$ ms. No breather forms.
}
\label{f:aas_MI-comparison}
\end{figure}

	First, using the (properly rescaled) eigenmode from the bottom edge of the highest band as an initial condition, we observed deviations from the optical eigenmode which were reinforced by the nonlinearity, leading to the eventual breakup of the waveform. As a result of this process, called modulational instability, a breather localized near the center of the lattice was formed.
	This is shown in the top panel of Fig. \ref{f:aas_MI-comparison}, which illustrates just how robust these breathers are.
	A power spectral density (PSD) analysis of these breathers revealed that they did indeed acquire frequencies in the upper band-gap; in this particular case, the dynamically resulting breather had a frequency of about 8.23 kHz~\cite{footnote}.
	The main conclusion from this and other similar evolution runs that we performed is that the top-band-edge breathers are prone to the manifestation of the modulational instability in ways similar to their lower gap analog in the dimer case experiments and analysis of Ref. \cite{Boechler10}.

	Since dissipation is ubiquitous in experimental setups~\cite{ricardo}, we then added linear on-site damping to each bead in order to determine whether dissipation would have any effect on the formation of these breathers due to modulational instability. 
	With this dissipation term, the equations of motion are modified to be
\begin{eqnarray} % Equations of motion with dissipation
m_n\ddot{u}_n &=&A_{n-1,n}[\delta_{n-1,n}-(u_n-u_{n-1})]_+^{3/2}
\nonumber 
\\
&-& A_{n,n+1}[\delta_{n,n+1}-(u_{n+1}-u_n)]_+^{3/2}\nonumber\\
&-& \frac{m_n\dot{u}_n}{\tau},
\label{e:dissipation}
\end{eqnarray}
where $\tau=2$ ms was chosen to best simulate the effects of dissipation in an experimental setting~\cite{georgenick}.
	The experiments above involving the linear mode as an initial condition were redone with dissipation turned on, and the results were strikingly different.
	Instead of observing modulational instability and the formation of a breather, the oscillations simply decayed exponentially.
	It appears that the timescales over which the solutions decay are too short to observe modulational instability.
	This comparison of the dissipative and Hamiltonian case can be seen between the bottom and top panels of Fig. \ref{f:aas_MI-comparison}.
	It is worth noting that in other simulations with much weaker dissipation, we did observe the formation of breathers, which were, in fact, more robust than those of the purely Hamiltonian case.
	This may become especially relevant if experimental techniques are devised that decrease the effects of dissipation. 
	See also below for a detailed discussion on the effects of dissipation for an experimentally relevant technique of producing such discrete breathers.

%%%%%%%%%%%%%%%%%%%%%%%%%%%%%%%%%%%%%%%%%%%
% Figure % A-A-S Modulational Instability %
%%%%%%%%%%%%%%%%%%%%%%%%%%%%%%%%%%%%%%%%%%%
%%%%%%%%%%%%%%%%%%%%%%%%%%%%%%%%%%%%%
%%%%%%%%%%%%%%%%%%%%%%%%%%%%%%%%%%%%%

\begin{figure}
\includegraphics[width=.45\textwidth]{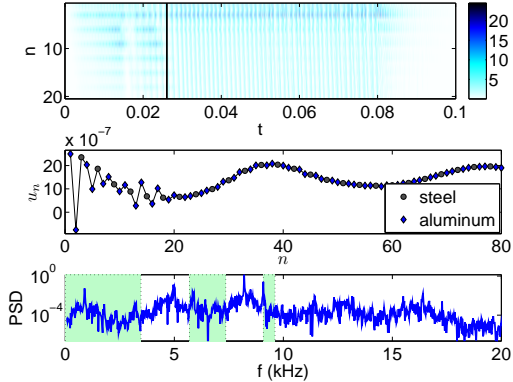}
\caption{
	(Color online) \emph{Top:} The velocities of an experiment with an actuator having frequency right at the lower edge of the third band ($f_d=9.084$ kHz), amplitude $9\times 10^{-7}$m, and dissipation (with $\tau=2$ ms). 
	\emph{Middle:} The profile of the chain at the time indicated by the vertical bar in the upper panel.
	This profile corresponds, at least qualitatively with both the surface breather and bulk breather of the same frequency.
	\emph{Bottom:} The PSD of the dynamical force experienced by bead 3.
	The linear bands are colored in so that the gaps are visible. 
	There is clear evidence of a frequency within the linear band gap, and the breather has a frequency of approximately 8.20 kHz.
}
\label{f:aas_actuator_MI}
\end{figure}

	Lastly, we investigated the possibility of nucleating one of these higher gap breathers using an actuator, a technique that has been experimentally employed in the dimer case in Ref. \cite{Boechler10}.
	Rendering the model more realistic, by keeping the dissipative term, we added a periodic (in time) drive beside the first bead in the chain (recall that previously, the chain had free boundary conditions). 
	We set this ``actuator,'' a harmonically oscillating bead at the left ($n=0$) boundary, to ramp up linearly to its full amplitude, then remain on for a period of time, and finally shut off, reverting to a fixed boundary.
\begin{equation}
u_0=\begin{cases}
a\left(\frac{t}{T_r}\right)^c\sin(\omega_d t) &\text{for }0\leq t\leq T_r\\
a\sin(\omega_d t)& \text{for }T_r<t\leq T_d\\
0&\text{for }t> T_d
\end{cases}
\end{equation}
where $a$ is the actuator amplitude, $T_r$ is the ramp-up time, $c=1$ for linear ramping, $\omega_d$ is the actuator frequency, and $T_d$ is the actuator duration.
	The accompanying intuition here is that with the actuator pumping at a frequency right at the bottom edge of the upper band, it is possible to excite, via modulational instability, the formation of a breather somewhere in the chain.
	With extensive experimentation, we found that the only breathers formed were on the surface of the chain, but as we mentioned above, these surface breathers do have similar profiles to those of the bulk breathers.
	Figure \ref{f:aas_actuator_MI} shows the results of one such experiment.
	Here, the actuator ramp time was 20 ms, the maximum amplitude of the actuator was $9\times 10^{-7}$ m, and the actuator was allowed to run for 80 ms before being shut off.
	A breather was formed right near the surface, which extended about two unit cells into the chain. 
	This breather had a frequency of about 8.20 kHz, which is well into the upper band gap.
	The profile of this breather is the same, at least qualitatively, with the corresponding exact solution found by the continuation at this frequency.

%%%%%%%%%%%%%%%%%%%%%%%%%%%%%%%%%%%%%%%%%%%%%
% Figure %    A-A-S Actuator Simulation     %
%%%%%%%%%%%%%%%%%%%%%%%%%%%%%%%%%%%%%%%%%%%%%
%%%%%%%%%%%%%%%%%%%%%%%%%%%%%%%%%%%%%%%%%%%%%
%%%%%%%%%%%%%%%%%%%%%%%%%%%%%%%%%%%%%%%%%%%%%

\subsection{Comparison to Other Trimers}

\begin{figure}
\includegraphics[width=.45\textwidth]{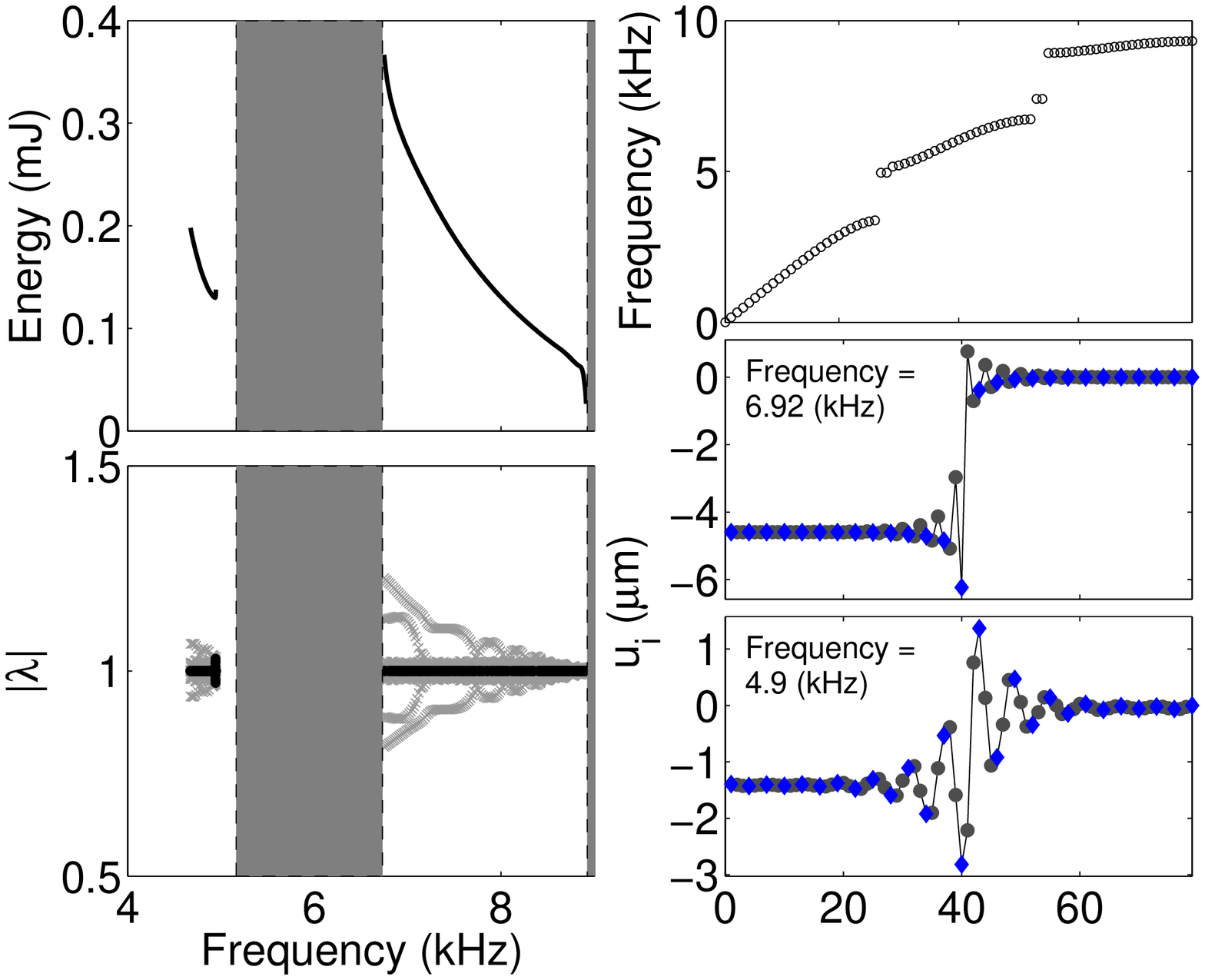}
\caption{
	(Color online) The results for the continuation of breathers in an `aluminum-steel-steel' chain. \emph{Top left:} The energy of each breather solution is plotted as a function of frequency. 
	\emph{Bottom left:} The complex modulus of each Floquet multiplier is shown as a function of its frequency.
	\emph{Top right:} The linear spectrum of the a-s-s chain is shown.
	\emph{Bottom right:} Two sample profiles of breathers in this chain are shown, one from each gap.
}\label{f:ass_continuation}
\end{figure}

\begin{figure}
\includegraphics[width=.45\textwidth]{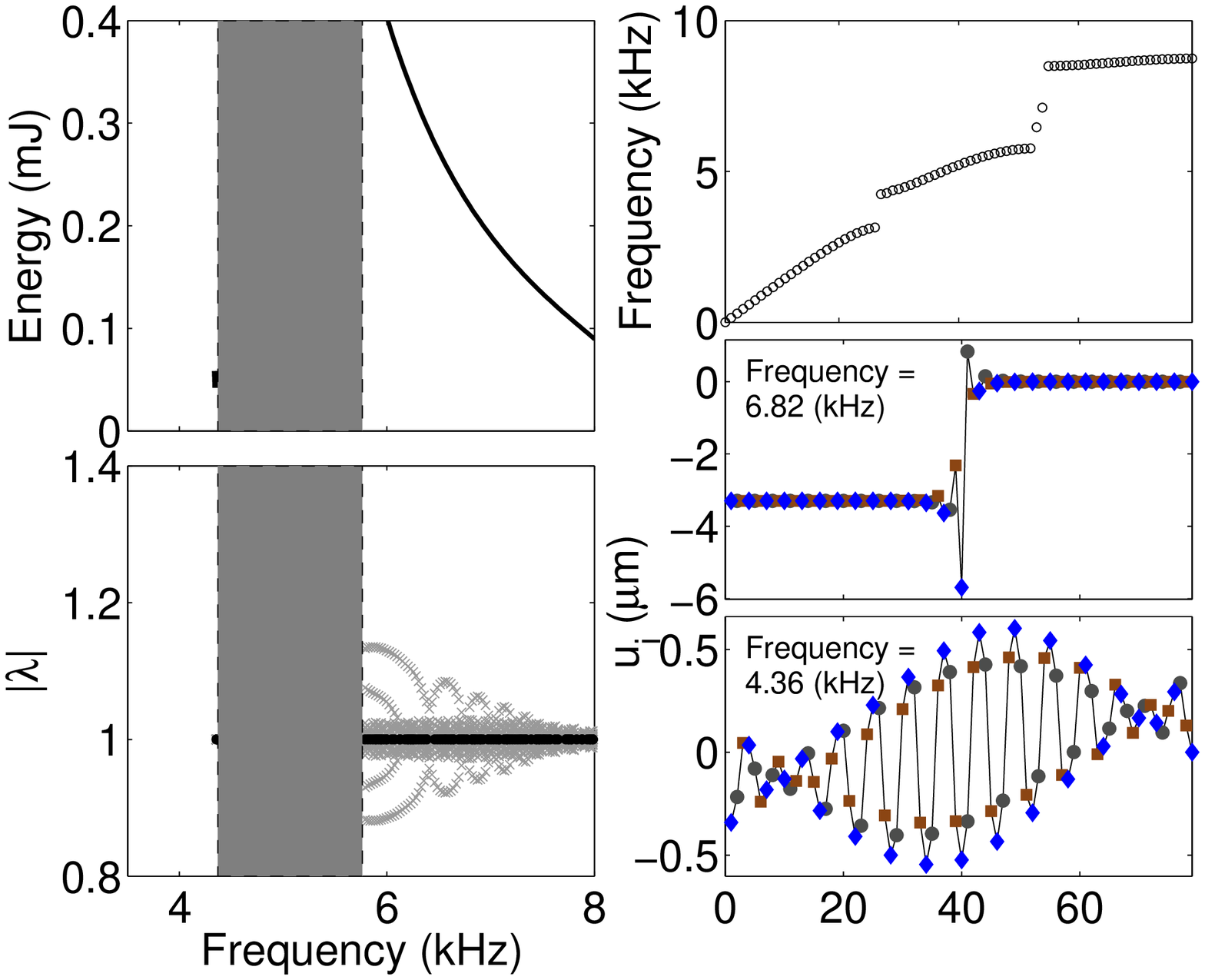}
\caption{
	(Color online) The results for the continuation of breathers in an `aluminum-steel-brass' chain. \emph{Top left:} the energy of each breather solution is plotted as a function of frequency. 
	\emph{Bottom left:} the complex modulus of each Floquet multiplier is shown as a function of its frequency.
	\emph{Top right:} the linear spectrum of the a-s-b chain is shown.
	\emph{Bottom right:} Two sample profiles of breathers in this chain are shown, one from each gap.
}\label{f:asb_continuation}
\end{figure}

	In this subsection the results of the a-a-s chain are compared to the results for two other trimers: one consisting of an `aluminum-steel-steel' (a-s-s) pattern, and the other consisting of an `aluminum-steel-brass' (a-s-b) pattern. 
	In the results of each of these continuations, shown in Figs. \ref{f:ass_continuation} and \ref{f:asb_continuation}, we see that for each of these configurations, the properties of the breathers are quite similar.
	More specifically, we identify as before three pass bands and between them two finite gaps (as well as a semi-infinite gap above the highest band) in the linear spectrum. 
	It is within these finite gaps that we focus on, regarding the potential existence of discrete breathers.
The lower gap has breathers that are mainly unstable, and not very localized, while the upper gap has breathers that are only weakly unstable, and are highly localized.
	The main difference between these two configurations and the previous (s-a-a) configuration is the absence of any real instability in the breathers of the higher gap, through a pair of Floquet multipliers exiting from unity onto the real axis,.i.e., the breather energy is found to have a monotonic dependence on the corresponding frequency. 
	Additionally, that critical frequency $f_c$ at which resonances with the second harmonic appear, due to the differences in the linear spectrum, is shifted in both cases.
	In particular, for the a-s-b chain, $f_c$ is shifted almost to the bottom edge of the second band, which prevents the numerical continuation from finding more than a small segment of breathers.
	Nevertheless, we highlight that all the principal characteristics of both the linear spectrum and of its gaps being conducive to the formation of the nonlinear waveforms of interest (with the same general structural features as before) are shared between these alternative configurations and the originally examined a-a-s one, attesting to the generic
nature of our conclusions.

\subsection{The Role of Dissipation}

\begin{figure}
\includegraphics[width=.45\textwidth]{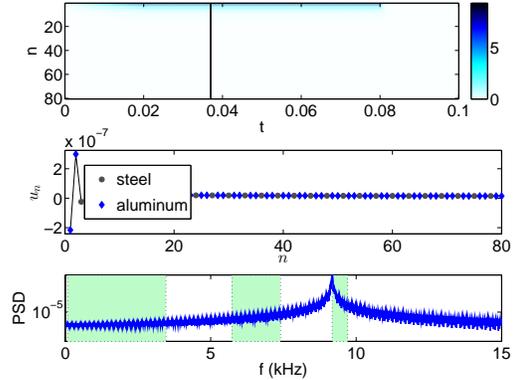}
\caption{
	(Color online) The diagnostics shown here are the same as in Fig.~\ref{f:aas_actuator_MI} but now for a dissipation which is 10 times larger, i.e., $\tau=0.2$ ms.
}\label{f:diss_02}
\end{figure}

\begin{figure}
\includegraphics[width=.45\textwidth]{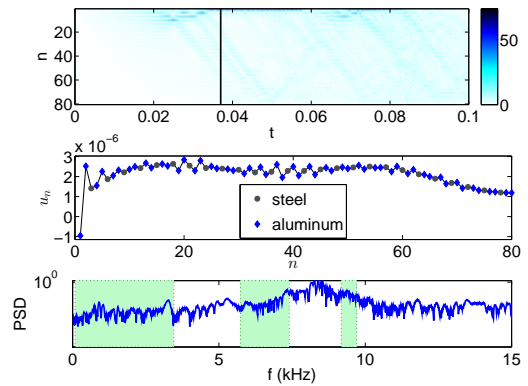}
\caption{
	(Color online) The diagnostics shown here are the same as in Fig.~\ref{f:aas_actuator_MI} but now for a dissipation which is 10 times smaller, i.e., $\tau=20$ ms in this case.
}\label{f:diss_20}
\end{figure}

	In this subsection, we examine the role of dissipation on the results associated with the actuation of the chain at its boundaries. 
	In particular, we perform a series of experiments which are identical to those of Fig.~\ref{f:aas_actuator_MI}, but now the dissipation is made either 10 times larger, i.e., $\tau=0.2$ms, as shown in Fig.~\ref{f:diss_02}, or it is 10 times smaller, i.e., $\tau=20$ms, as presented in Fig.~\ref{f:diss_20}. 
	In the former case, it is clear that the dissipation completely overwhelms any potential nonlinear structure formation, coming into play at a very early time scale. 
	It is for that reason that the dynamics of the fifth site is purely oscillatory, and the relevant oscillation frequency, as illustrated in the corresponding power spectral density, is found to be the driving frequency at the bottom edge of the third band. 
	On the contrary, and in a way more reminiscent of the dynamics of Fig.~\ref{f:aas_actuator_MI}, the dynamics of the case with $\tau=20$ ms is more conducive to the emergence of additional eigenfrequencies through the modulational instability mechanism highlighted previously. 
	It is for that reason that the evolution of the force at the fifth site appears to involve multiple frequencies, as is confirmed also by the corresponding power spectral density.
	Furthermore, as expected, in addition to the driving frequency, there exists a maximally excited frequency within the second gap which is associated with the discrete surface breather emerging in the dynamics of the top panel.

	In conclusion, for the range of dissipations that have been recently used to match experimental results~\cite{georgenick}, as well as for weaker dissipations, the waveforms identified herein are expected to persist. 
	On the other hand, for the case of materials with significantly more dissipative dynamics (and possibly effects of plasticity that additionally drain
the energy pumped into the system), these coherent structures cannot develop.

%%%%%%%%%%%%%%%%%%%%%%%%%%%%%%%%%%%%%%%%
% Section %        A-A-A-S Chain       %
%%%%%%%%%%%%%%%%%%%%%%%%%%%%%%%%%%%%%%%%
\section{A-A-A-S Chain}

\subsection{Linear Spectrum}

\begin{figure}
\includegraphics[width=.45\textwidth]{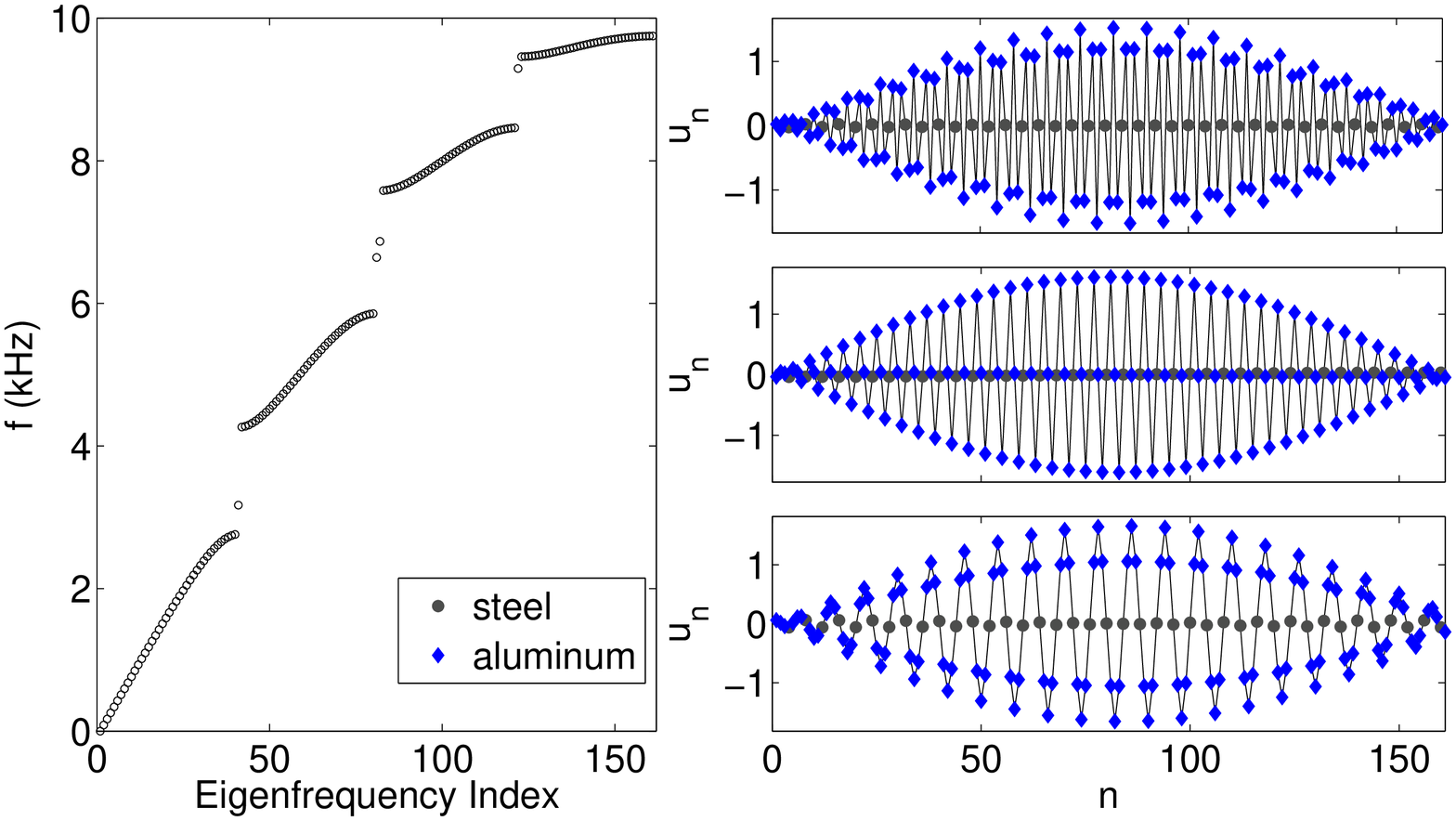}
\caption{
	(Color online) The linear spectrum for a chain with period 4 (aluminum-aluminum-aluminum-steel). The spectrum features three finite gaps separating the four existing pass bands. 
	Within the gaps, the (free) boundary conditions induce a number of
localized surface modes (at the chain boundaries).
}
\label{f:aaas_linear_spectrum}
\end{figure}

	In this section we present similar results to the above a-a-s chain, 
but for  a chain with periodicity involving $P=4$ sites, with each unit cell following an a-a-a-s pattern. 
	Similarly to the a-a-s case, we start by illustrating in Fig. \ref{f:aaas_linear_spectrum} the linear spectrum of a chain with 160 beads, which has four bands and three finite gaps. 
	Examples of the linear modes at the bottom of each of the bands (second, third, and fourth, respectively) are shown in the right panel of Fig. \ref{f:aaas_linear_spectrum}.
	It can be seen that at the bottom of the second band, the mode involves an in-phase excitation of the aluminum beads (while the steel ones are nearly stationary).
	At the bottom of the third band every other bead (one steel and one aluminum) are essentially stationary, while the remaining two aluminum beads that straddle the stationary one are out of phase with respect to each other. 
	Finally at the bottom of  fourth band, each aluminum bead is essentially out of phase with its neighboring one (while the heavier steel beads remain stationary).

\subsection{Existence and Stability of Discrete Breathers}

\begin{figure}
\includegraphics[width=.45\textwidth]{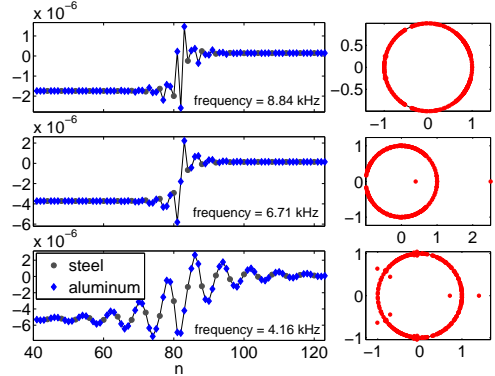}
\caption{
	(Color online) \emph{Top:} A sample breather profile from the third band gap, along with its Floquet multipliers.
	\emph{Middle:} A sample breather profile from the middle (finite)
band gap.
	\emph{Bottom:} A sample breather profile from the lowest band gap.
}
\label{f:aaas_breather_profiles}
\end{figure}

	Once again, a continuation through each gap was performed, starting from the corresponding band edge modes and, accordingly, three families of breathers were discovered. 
	A sample breather from each gap is shown in Fig. \ref{f:aaas_breather_profiles}.
	It can be seen that each of the breathers preserves a structure reminiscent of the corresponding linear mode from which it stemmed, while forming an increasing displacement step due to the asymmetry of the potential (as discussed earlier in Ref. \cite{theo10}). 
	As before, the breathers acquire an increasing degree of localization as the frequency gets deeper into the corresponding band gap.

\begin{figure}
\includegraphics[width=.45\textwidth]{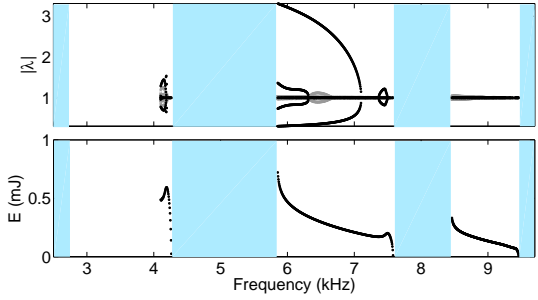}
\caption{
(Color online) \emph{Top:} The magnitude of each Floquet multiplier is shown as a function of its frequency.
	Purely real Floquet multipliers are shown in black, while Floquet multipliers with nonzero imaginary part are shown in gray.
 \emph{Bottom:} The energy of each solution as a function of frequency.
 The horizontal scale is the same for both panels.
 	In both plots, the frequency ranges of the linear bands are 
shaded for clarity.
}
\label{f:aaas_energy_floq}
\end{figure}

	Finally, the linear stability of each breather was computed via Floquet analysis, and the magnitude of the Floquet multipliers was plotted as a function of frequency in Fig. \ref{f:aaas_energy_floq}.
	Remarkably, once again breathers in the uppermost finite gap are the most robust among the three families. 
	In particular, they feature no real instability whatsoever and, in fact, have an interval of complete stability, while their oscillatory instabilities are only very weak when present.
	On the other hand, the middle gap breathers are highly unstable, featuring a strong real instability in a large fraction of the corresponding band gap (and even a second real instability close to the top of the second band). 
	Nevertheless, they also possess an interval of stability in the immediate vicinity of of the bottom of the third band; there is also a mild real instability in that neighborhood associated with an energy slope change; see Fig. \ref{f:aaas_energy_floq} and cf. Ref. \cite{theo10}.
	Finally, within the lower gap, the discrete granular breathers are so spatially extended that boundaries seem to interfere with our numerical continuation demanding a very large domain to accurately perform the relevant computation. 
	Nevertheless, these states also appear to be subject to strong real instabilities. 
	Thus, once again the top finite gap breathers appear to be the most robust nonlinear entities of the system.

\begin{figure}
\includegraphics[width=.45\textwidth]{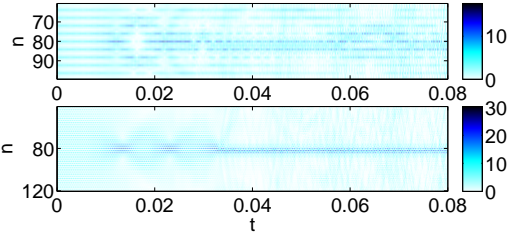}
\caption{
	(Color online) The results of a  numerical 
experiment of the modulational instability of a linear mode at the bottom edge of the third (\emph{top}) and second (\emph{bottom}) optical band.
	In each case a breather was formed, which was localized near the center of the lattice.
	These breathers had frequencies of approximately 9.12 kHz and 7 kHz, respectively, in the second and third finite band gaps.
}\label{f:aaas_MI}
\end{figure}

\subsection{Dynamics and Modulational Instability}

	We also investigated the potential for exciting discrete breathers in the 3:1 chain.
	Once again, we used a rescaled linear mode from the bottom edge of each optical band as an initial condition in our numerical integrator, and observed that modulational instability did indeed form a breather localized near the center of the chain. 
	Figure \ref{f:aaas_MI} shows the results of these experiments.
	A PSD analysis shows that the frequency of each breather is located well into the corresponding band gaps.
	
\begin{figure}
\includegraphics[width=.45\textwidth]{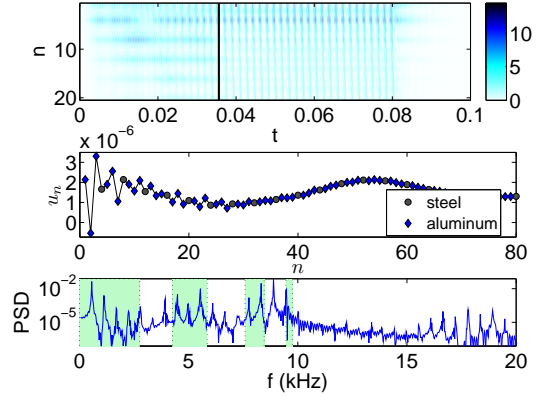}
\caption{
	(Color online) \emph{Top:} The velocities of a numerical 
experiment with an actuator having frequency at the bottom of the fourth band (9.45 kHz), amplitude $5\times 1.5^{-6}$, and dissipation (with $\tau=2$). The chain in this case had 159 beads in order for boundary reflection effects to be lessened, and a clearer picture obtained.
	\emph{Bottom:} The PSD of the force time series of bead 5 is shown.
	The linear bands are colored in so that the gaps are visible. 
	There is clear evidence of a frequency within the top band gap, and the breather has a frequency of approximately 8.89 kHz.
}
\label{f:aaas_actuator_MI}
\end{figure}

\begin{figure}
\includegraphics[width=.45\textwidth]{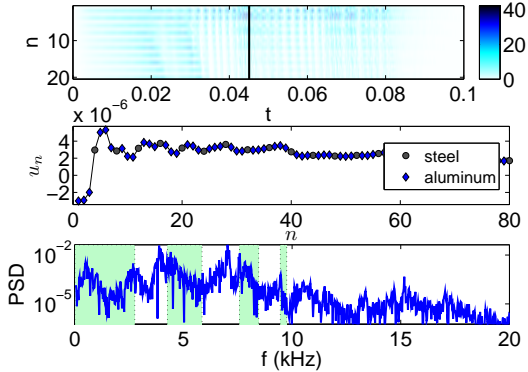}
\caption{
(Color online) \emph{Top:} The velocities of an experiment with an actuator having frequency at the bottom of the third band (7.59 kHz), amplitude $8\times 10^{-7}$, and dissipation (with $\tau=2$).
	\emph{Bottom:} The PSD of the force time series of bead 5 is shown.
	The linear bands are colored in so that the gaps are visible. 
	There is clear evidence of a frequency within the middle finite band gap, and the breather has a frequency of approximately 7.03 kHz.
}
\label{f:aaas_actuator-mid-gap}
\end{figure}

	In order to excite the relevant modes, we have once again resorted to the actuation of the band edge of (the bottom of) the different pass bands, in line with our description in the previous section (we similarly use a ramp and eventually cut off the driving of the chain boundary). 
	We have also included dissipation in the model to render our results closer to the experimentally anticipated ones.
	Our findings are shown in Figs. \ref{f:aaas_actuator_MI} and \ref{f:aaas_actuator-mid-gap}, respectively, for the top and middle finite gaps. 
	The conclusion in both cases is, in fact, similar, namely the drive of the mode at the bottom of the optical band edge induces a modulationally unstable evolution, in turn, leading to localization within the respective band gap.
	As a result the former figure features a breather of frequency 8.89 kHz, as a result of the dynamical evolution, while the latter one similarly possesses a localized mode of frequency 7.03KHz within the middle gap. 
	These results indicate that the experimental realization of such higher gap breathers should certainly be within reach with currently accessible experimental settings.

\section{Conclusions and Future Challenges}

	In the present work, we have extended earlier considerations focusing on precompressed elastic dimers (featuring nonlinear discrete breather modes within their single finite spectral gap).
	In particular, we have considered the linear and associated nonlinear properties of, arguably, one of the simplest forms of trimer and of quadrimer chains, namely a 2:1 aluminum-steel chain and a 3:1 such chain. 
	We have illustrated the presence of 2 finite band gaps in the first case and of three such in the second. 
	Within these bandgaps, we have been able to identify (starting from the bottom edge of the upper band in each case) a sequence of gap breather localized modes with increased localization properties, as the frequency of their temporal oscillation proceeded deeper into the linear gap.
	We have also showcased the highly nontrivial feature that the states of the higher gaps appear, in fact, to be more robust than the ones of the lower gaps, presumably a feature that should be attributed to their out-of-phase spatial structure. 
	Finally, we have also demonstrated the spontaneous emergence of such states, as a result not only of the modulational instability of the extended modes of the bottom edge of the optical bands of the linear spectrum, but also as a result of the experimentally realizable scenario of the system being driven at such band edge frequencies.

	We believe that these findings pave the way towards the experimental realization and observation of such states and more generally towards a systematic classification of gap breather modes in elastic systems with multiple band gaps. 
	On the other hand, from the theoretical perspective, it would be of particular interest to generalize the present considerations not only to alternative one-dimensional heterogeneous configurations, but perhaps more appealingly to the substantially more challenging setting of two-dimensional granular crystals. 
	Understanding the bandgaps of such two-dimensional chains and the associated intrinsically localized modes thereof would be a theme of particular relevance for theoretical and experimental studies alike.

\vspace{5mm}

\begin{acknowledgments}
 PGK gratefully
acknowledges support from NSF-CMMI-1000337 and from the A.S. Onassis
Public Benefit Foundation through Grant No. RZG 003/2010-2011. 
Both authors are greatly indebted to Dr. George Theocharis, Dr. Nick Boechler
and Prof. Chiara Daraio of CalTech, for sharing numerous details and
insights about their experimental granular chain setup and its modeling.
\end{acknowledgments}

\appendix
\section{Appendix: Poincar\'{e} Continuation}
\label{a:poincare}
	In order to find periodic solutions to this system, we are searching for solutions $u$ so that $u_n(0)=u_n(T)$ where $T$ is the fixed period of the solution. 
	This gives us a Poincar\'{e} map, $P(u_0)=u(0;u_0)-u(T;u_0)$, where $u_0$ is the initial condition, and $u(t;u_0)$ is a solution to system of ODEs at time $t$ with initial condition $u_0$. 
	A periodic solution with period $T$ (frequency $1/T$) will be a root of $P$. 
	Since any point on a periodic solution will also be a root, we add the additional constraints that $\dot{u}_n=0$ in order to ``pin" the solution to a point in the domain of the Poincar\'{e} map. 
	To calculate the numerically exact (up to a set relative error, usually $10^{-10}$) profile of each periodic solution, we used a Newton-Raphson solver to calculate the fixed-points of $P$.
	The Jacobian for the Newton's method is given by $I-V(T)$, where $V(T)$ is the monodromy matrix, which we obtain by integrating the variational equation
	$$ \frac{d}{dt}V=J(u,\dot{u})V,$$
simultaneously with the equations of motion \eqref{e:motion}, where
$J$ is the Jacobian of these nonlinear equations, and $V(0)=I$ (i.e.,
the
intial condition of the monodromy matrix is the identity matrix).
	The eigenvalues of the monodromy matrix also give us the Floquet multipliers of the solution, which convey information about the linear stability of the obtained periodic orbit. 
	Floquet multipliers outside the unit circle give rise to instabilities, while Floquet multipliers on the unit circle lead to marginal stability, and Floquet multipliers purely inside the unit circle give rise to asymptotic stability.

	Once we have our numerically exact solution $u$, which is hopefully a localized DB, we then change the frequency (and thus the period) of our desired solution, and using $u$ as an initial guess, we again use the Newton solver to find another exact solution.
	An entire family of solutions can then be found by `continuing' this solution through the frequency gap.
	For a review of the many methods of finding DBs, see Ref. \cite{flach09} and the references therein.
	For more details about numerical continuation, see Ref. \cite{doedel91}.


\begin{thebibliography}{99}
\bibitem{footnote}A particularly interesting feature associated with this observation concerns the mechanism of selection of a particular frequency for the resulting breather.
	Apparently, different mode amplitudes will lead to breathers of different frequencies, yet the associated selection/correspondence mechanism is not well-understood.
\end{thebibliography}
\end{document}